\documentclass[english,aps, twocolumn]{revtex4-1}
\usepackage[T1]{fontenc}
\usepackage[latin9]{inputenc}
\usepackage{color}
\usepackage{babel}
\usepackage{amsmath}
\usepackage{amssymb}
\usepackage{graphicx}
\usepackage[unicode=true,pdfusetitle,
 bookmarks=true,bookmarksnumbered=false,bookmarksopen=false,
 breaklinks=false,pdfborder={0 0 0},pdfborderstyle={},backref=false,colorlinks=true]
 {hyperref}
\hypersetup{
 linkcolor=darkblue, citecolor=darkblue, urlcolor=darkblue}

\makeatletter

\usepackage{color}
\usepackage{babel}
\usepackage{breakurl}

\renewcommand{\fnum@figure}{FIG.~\thefigure}

\definecolor{darkblue}{rgb}{0.0, 0.0, 0.55}
\usepackage{feynmp-auto}

\@ifundefined{showcaptionsetup}{}{
 \PassOptionsToPackage{caption=false}{subfig}}

\usepackage[caption=false]{subfig} 

\@ifundefined{showcaptionsetup}{}{%
 \PassOptionsToPackage{caption=false}{subfig}}
\usepackage{subfig}
\makeatother

\begin{document}

\title{Inverse Bremsstrahlung current drive}

\author{Vadim R. Munirov}
\email{vmunirov@pppl.gov}

\selectlanguage{english}%

\affiliation{Princeton Plasma Physics Laboratory, Princeton University, Princeton,
New Jersey 08543, USA}

\affiliation{Department of Astrophysical Sciences, Princeton University, Princeton,
New Jersey 08540, USA}

\author{Nathaniel J. Fisch}

\affiliation{Princeton Plasma Physics Laboratory, Princeton University, Princeton,
New Jersey 08543, USA}

\affiliation{Department of Astrophysical Sciences, Princeton University, Princeton,
New Jersey 08540, USA}

\date{22 August 2017}
\begin{abstract}
The generation of the plasma current resulting from Bremsstrahlung
absorption is considered. It is shown that the electric current is
higher than the naive estimates assuming that electrons absorb only
the photon momentum and using the Spitzer conductivity would suggest.
The current enhancement is in part because electrons get the recoil
momentum from the Coulomb field of ions during the absorption and
in part because the electromagnetic power is absorbed asymmetrically
within the electron velocity distribution space. 
\end{abstract}
\maketitle

\section{Introduction}

In the presence of external electromagnetic field colliding electrons
and ions absorb the incoming radiation through the process known as
inverse Bremsstrahlung. In Bremsstrahlung absorption, the electron
receives additional recoil momentum from the ion besides the momentum
of the photon. Therefore, plasma electrons absorb more than just the
photon momentum from the incoming radiation. The generated current
is then larger than one would get by assuming that electrons absorb
just the photon momentum. It was shown in \cite{Pashinin1978} that
this increase in current is equal to $8/5$. 

However, the recoil is not the only mechanism that will increase the
current. Plasma electrons absorb the radiation asymmetrically in velocity
space; specifically, electrons co-moving with the incoming photons
will absorb slightly more power than electrons going in the opposite
direction. Even in the absence of net momentum absorption, this asymmetric
absorption in power can lead to current drive. This is because the
collision frequency in plasma is speed dependent. Thus, upon absorbing
energy electrons going in the direction of the incoming radiation
will experience less resistance from the plasma than electrons going
in the opposite direction resulting in current. This is called the
asymmetric resistivity current drive effect and is mostly known with
respect to cyclotron absorption used to drive toroidal current in
tokamaks \cite{FischBoozer1980,Fisch1987}. Moreover, even without
the asymmetric resistivity effect the fluid approximation is less
precise in considering current generation as opposed to momentum input,
because it assumes that all electrons get equal push in the same direction,
which is not the case for Bremsstrahlung absorption. In fact, the
ability of electrons to retain current is sensitive to both its location
in velocity space and the direction in which it is being pushed.

In this paper we rederive the result for the momentum absorption rate
and calculate the additional increase in current due to the current
drive effect. To derive the current drive effect, it will be necessary
to consider in detail how exactly the momentum is absorbed within
the electron velocity space. To do this we use the formalism developed
by Tsytovich \cite{Tsytovich1992,Tsytovich1995,Tsytovich1996}.

\section{Probability of Bremsstrahlung}

Consider Bremsstrahlung absorption for particles $\alpha$ (electrons)
due to the Coulomb collisions with much heavier particles $\beta$
(ions). To satisfy the conservation laws of momentum and energy, in
each act of the Bremsstrahlung absorption some recoil momentum must
be transferred from the electron to ions. We can write down the momentum
balance during inverse Bremsstrahlung as follows:

\begin{align}
\mathbf{p}_{\alpha}^{\prime} & =\mathbf{p}_{\alpha}+\hbar\mathbf{k}-\hbar\mathbf{q},\\
\mathbf{p}_{\beta}^{\prime} & =\mathbf{p}_{\beta}+\hbar\mathbf{q},
\end{align}

\noindent where the primed values correspond to the quantities after
the absorption, $\mathbf{k}$ is the wave vector of the photon, and
$\mathbf{q}$ is the recoil wave vector transferred from the electron
to the ion. The conservation of energy is

\begin{equation}
\varepsilon_{\mathbf{p}_{\alpha}}^{\alpha}+\varepsilon_{\mathbf{p}_{\beta}}^{\beta}+\hbar\omega_{\mathbf{k}}=\varepsilon_{\mathbf{p}_{\alpha}+\hbar\mathbf{k}-\hbar\mathbf{q}}^{\alpha}+\varepsilon_{\mathbf{p}_{\beta}+\hbar\mathbf{q}}^{\beta}.
\end{equation}

Here, we will use the diffusion approximation, when $\hbar\mathbf{k}$,
$\hbar\mathbf{q}$ are small in comparison with the particle momentum
$(\hbar\mathbf{k},\:\hbar\mathbf{q}\ll\mathbf{p}_{\alpha})$. In this
approximation, the energy conservation is simplified to

\begin{equation}
\omega_{\mathbf{k}}=\left(\mathbf{k}-\mathbf{q}\right)\mathbf{v}_{\alpha}+\mathbf{q}\mathbf{v}_{\beta}.\label{eq:cond_1}
\end{equation}

Now consider the direct process of spontaneous Bremsstrahlung emission.
The momentum balance can be written as:

\begin{align}
\mathbf{p}_{\alpha}^{\prime} & =\mathbf{p}_{\alpha}-\hbar\mathbf{k}+\hbar\mathbf{q},\\
\mathbf{p}_{\beta}^{\prime} & =\mathbf{p}_{\beta}-\hbar\mathbf{q}.
\end{align}

With such a definition of the recoil momentum\textbf{ $\mathbf{q}$}
(notice different signs in the definition of \textbf{q} for emission
and absorption), the energy conservation yields the same relationship
between velocities of the particles and parameters of the photon as
for the inverse process (Eq.~(\ref{eq:cond_1})).

\begin{figure}
\subfloat[\label{fig01_a}]{\includegraphics[width=0.5\columnwidth]{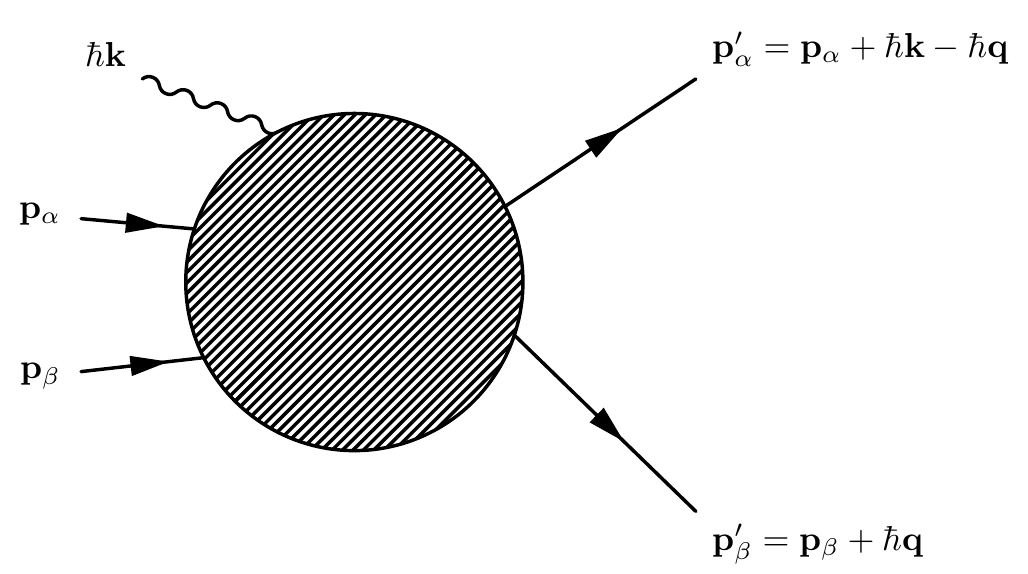}

}\subfloat[\label{fig01_b}]{\includegraphics[width=0.5\columnwidth]{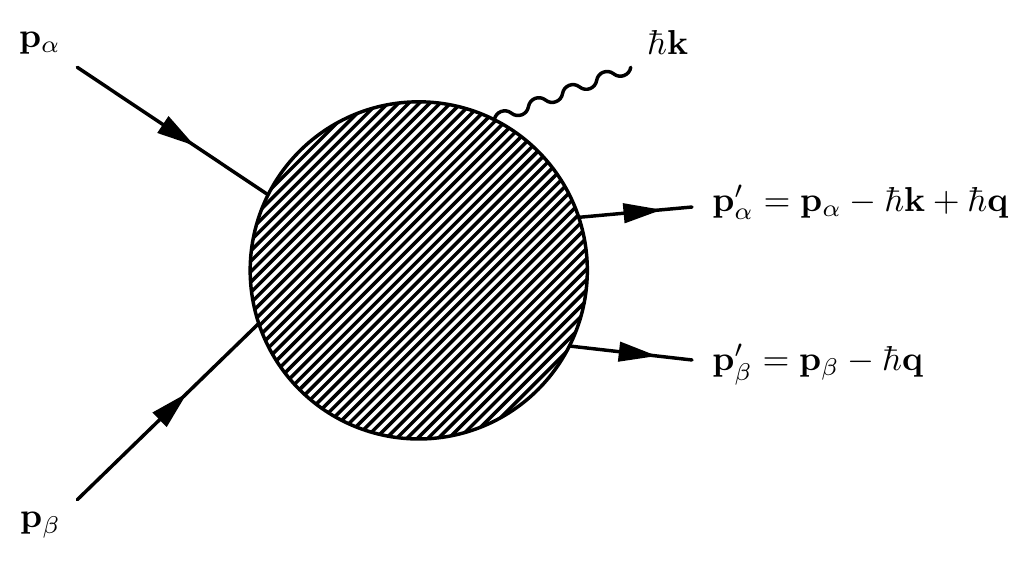}

}

\caption{\label{fig01}Schematic diagram of Bremsstrahlung absorption (a) and
emission (b).}
\end{figure}

A schematic diagram of the two processes is shown in Fig.~\ref{fig01}.
Essentially inverse Bremsstrahlung can be considered as Compton scattering,
by the incoming electron, of the incoming photon $\mathbf{k}$ into
the virtual photon of the Coulomb field $\mathbf{q}$ (see Fig.~\ref{fig01_a}),
while the Bremsstrahlung emission can be considered as Compton scattering
of the virtual photons of the Coulomb field on the incoming electron
(see Fig.~\ref{fig01_b}).

It is clear, that due to time reversal symmetry, the transition probability
of the inverse and direct processes are related to each other:

\begin{equation}
w_{\mathbf{p}_{\alpha},\mathbf{p}_{\beta}}^{IBr}\left(\mathbf{k},\mathbf{q}\right)=w_{\mathbf{p}_{\alpha}+\hbar\mathbf{k}-\hbar\mathbf{q},\mathbf{p}_{\beta}+\hbar\mathbf{q}}^{Br}\left(\mathbf{k},\mathbf{q}\right).
\end{equation}

\noindent Here $w_{\mathbf{p}_{\alpha},\mathbf{p}_{\beta}}^{Br}\left(\mathbf{k},\mathbf{q}\right)$
and $w_{\mathbf{p}_{\alpha},\mathbf{p}_{\beta}}^{IBr}\left(\mathbf{k},\mathbf{q}\right)$
are the probabilities of spontaneous Bremsstrahlung emission and inverse
Bremsstrahlung per unit time within $d\mathbf{k}d\mathbf{q}$. Note
that these probabilities must contain condition (\ref{eq:cond_1})
as the argument of the delta function.

One must remember that, in the presence of external radiation, the
true absorption due to inverse Bremsstrahlung is always accompanied
by the process of stimulated emission. For example, for electromagnetic
waves ($\omega=kc$) and infinitely massive ions ($\mathbf{v}_{\beta}=0$),
condition (\ref{eq:cond_1}) implies that for inverse Bremsstrahlung
the change in the parallel momentum of the electron is approximately
$\hbar\omega/v$, while for stimulated Bremsstrahlung emission this
change is approximately $-\hbar\omega/v$. However, these two processes
do not completely compensate each other because their probabilities
are slightly different. 

More generally, the evolution of the distribution function $f_{\mathbf{p}_{\alpha}}^{\alpha}$
due to the processes of inverse Bremsstrahlung and stimulated Bremsstrahlung
emission is described by \cite{Tsytovich1992}

\begin{multline}
\frac{\partial f_{\mathbf{p}_{\alpha}}^{\alpha}}{\partial t}=-\int w_{\mathbf{p}_{\alpha},\mathbf{p}_{\beta}}^{IBr}\left(\mathbf{k},\mathbf{q}\right)f_{\mathbf{p}_{\alpha}}^{\alpha}f_{\mathbf{p}_{\beta}}^{\beta}n_{\mathbf{k}}d\mathbf{k}d\mathbf{q}d\mathbf{p}_{\beta}\\
+\int w_{\mathbf{p}_{\alpha}-\hbar\mathbf{k}+\hbar\mathbf{q},\mathbf{p}_{\beta}-\hbar\mathbf{q}}^{IBr}\left(\mathbf{k},\mathbf{q}\right)f_{\mathbf{p}_{\alpha}-\hbar\mathbf{k}+\hbar\mathbf{q}}^{\alpha}f_{\mathbf{p}_{\beta}-\hbar\mathbf{q}}^{\beta}n_{\mathbf{k}}d\mathbf{k}d\mathbf{q}d\mathbf{p}_{\beta}\\
-\int w_{\mathbf{p}_{\alpha},\mathbf{p}_{\beta}}^{Br}\left(\mathbf{k},\mathbf{q}\right)f_{\mathbf{p}_{\alpha}}^{\alpha}f_{\mathbf{p}_{\beta}}^{\beta}n_{\mathbf{k}}d\mathbf{k}d\mathbf{q}d\mathbf{p}_{\beta}\\
+\int w_{\mathbf{p}_{\alpha}+\hbar\mathbf{k}-\hbar\mathbf{q},\mathbf{p}_{\beta}+\hbar\mathbf{q}}^{Br}\left(\mathbf{k},\mathbf{q}\right)f_{\mathbf{p}_{\alpha}+\hbar\mathbf{k}-\hbar\mathbf{q}}^{\alpha}f_{\mathbf{p}_{\beta}+\hbar\mathbf{q}}^{\beta}n_{\mathbf{k}}d\mathbf{k}d\mathbf{q}d\mathbf{p}_{\beta}.
\end{multline}

Following Tsytovich \cite{Tsytovich1992,Tsytovich1995,Tsytovich1996},
after Taylor expansion for $\hbar\mathbf{k},\:\hbar\mathbf{q}\ll\mathbf{p}_{\alpha}$
we get the Fokker-Planck equation for the evolution of $f_{\mathbf{p}_{\alpha}}^{\alpha}$:

\begin{equation}
\frac{\partial f_{\mathbf{p}_{\alpha}}^{\alpha}}{\partial t}=\frac{\partial}{\partial\mathbf{p}_{\alpha}}\cdot\mathbf{S}_{\mathbf{p}_{\alpha}}=\frac{\partial}{\partial\mathbf{p}_{\alpha}}\cdot\left(\widehat{D}_{\alpha}\frac{\partial f_{\mathbf{p}_{\alpha}}^{\alpha}}{\partial\mathbf{p}_{\alpha}}+\mathbf{F}_{\alpha}f_{\mathbf{p}_{\alpha}}^{\alpha}\right),\label{eq:FP}
\end{equation}

\noindent where

\begin{equation}
\widehat{D}_{\alpha}=\int\hbar^{2}\left(\mathbf{k}-\mathbf{q}\right)\left(\mathbf{k}-\mathbf{q}\right)w_{\mathbf{p}_{\alpha},\mathbf{p}_{\beta}}^{Br}n_{\mathbf{k}}f_{\mathbf{p}_{\beta}}^{\beta}d\mathbf{k}d\mathbf{q}d\mathbf{p}_{\beta},
\end{equation}

\begin{equation}
\mathbf{F}_{\alpha}=\int\hbar^{2}\left(\mathbf{k}-\mathbf{q}\right)\left(\mathbf{q}\cdot\frac{\partial f_{\mathbf{p}_{\beta}}^{\beta}}{\partial\mathbf{p}_{\beta}}\right)w_{\mathbf{p}_{\alpha},\mathbf{p}_{\beta}}^{Br}n_{\mathbf{k}}d\mathbf{k}d\mathbf{q}d\mathbf{p}_{\beta}.
\end{equation}

\noindent The normalization is such that the density of particles
is $n_{\alpha}=\int f_{\mathbf{p}_{\alpha}}^{\alpha}d\mathbf{p}_{\alpha}=\int f_{\mathbf{v}_{\alpha}}^{\alpha}d\mathbf{v}_{\alpha},$
and the total number of photons per volume is $N_{ph}=\int n_{\mathbf{k}}d\mathbf{k}$,
and $n_{\mathbf{k}}$ is the number of photons within $d\mathbf{k}$.

The probability of spontaneous Bremsstrahlung emission for electromagnetic
waves $(\omega=kc)$ keeping terms of the order of $\mathbf{k}\mathbf{v}/\omega\sim v/c$
is given by \cite{Tsytovich1996}

\begin{multline}
w_{\mathbf{p}_{\alpha},\mathbf{p}_{\beta}}^{Br}\left(\mathbf{k},\mathbf{q}\right)=\frac{2e_{\alpha}^{4}e_{\beta}^{2}\delta\left[\omega_{\mathbf{k}}-\left(\mathbf{k}-\mathbf{q}\right)\mathbf{v}_{\alpha}-\mathbf{q}\mathbf{v}_{\beta}\right]}{\hbar\pi^{2}m_{\alpha}^{2}q^{4}\left(\omega_{\mathbf{k}}-\mathbf{k}\mathbf{v}_{\alpha}\right)^{2}\left.\frac{\partial\left(\varepsilon\omega^{2}\right)}{\partial\omega}\right|_{\omega=\omega_{\mathbf{k}}}\varepsilon_{\mathbf{q},\mathbf{q}\mathbf{v}_{\beta}}^{2}}\\
\times\left|\left[\mathbf{e}_{k}\times\mathbf{q}\right]+\frac{\mathbf{k}\mathbf{q}}{\omega_{\mathbf{k}}-\mathbf{k}\mathbf{v}_{\alpha}}\left[\mathbf{e}_{k}\times\mathbf{v}\right]\right|^{2}.\label{eq:prob}
\end{multline}

This expression is only correct for Bremsstrahlung ignoring the polarization
effects. By polarization effects we mean that the plasma environment
in which the electron finds itself is influenced by the presence of
the electron. This approximation is good for dilute plasma. In general,
the probability of Bremsstrahlung is proportional to $\left|\left[\mathbf{e}_{k}\times\left(\mathbf{M}^{\alpha}+\mathbf{M}^{\beta}+\mathbf{M}^{\alpha\beta}\right)\right]\right|^{2}$,
where $\mathbf{M}^{\alpha}$ is the emission due to oscillation of
$\alpha$ particles in the screened field of $\beta$ charges, $\mathbf{M}^{\beta}$
is the emission due to oscillation of $\beta$ particles in the screened
field of $\alpha$ charges, and $\mathbf{M}^{\alpha\beta}$ is the
emission due to oscillation of the polarization clouds around particles
$\alpha$ and $\beta$. While $\mathbf{M}^{\beta}$ is small due to
the high ion mass, the term $\mathbf{M}^{\alpha\beta}$ can be comparable
with $\mathbf{M}^{\alpha}$. Moreover, polarization effects may make
electron-electron and ion-ion collisions important as well. The polarization
effects are especially important for longitudinal waves, and must
be almost always taken into account for them (we consider only transverse
electromagnetic waves here) \cite{Tsytovich1992,Tsytovich1995,Tsytovich1996}.
In Eq.~(\ref{eq:prob}) the polarization effects are ignored and
only $\mathbf{M}^{\alpha}$ term is retained; this requires the plasma
to be tenuous enough. Another approximation used in Eq.~(\ref{eq:prob})
is non-relativistic velocities. In all subsequent calculations, we
also take unity dielectric function ($\varepsilon\approx1$), which
is a good approximation for tenuous plasma. We will also ignore plasma
dispersive effects and take $\omega_{\mathbf{k}}=\omega=kc$, and
assume an infinite ion mass and set $\mathbf{v}_{\beta}=0$, $\mathbf{v}_{\alpha}=\mathbf{v}$. 

\section{Momentum change}

In this section let us calculate the rate of momentum change for electrons
during Bremsstrahlung absorption.

\noindent From Eq.~(\ref{eq:FP}) we can calculate the rate of momentum
absorption due to Bremsstrahlung as:

\begin{equation}
\frac{d\mathbf{p}_{V}^{\alpha}}{dt}=-\int\mathbf{S}_{\mathbf{p}_{\alpha}}d\mathbf{p}_{\alpha},\label{eq:S_p}
\end{equation}

\noindent so $-\mathbf{S}_{\mathbf{p}_{\alpha}}$ has the meaning
of the rate of momentum absorption per $d\mathbf{p}_{\alpha}$ by
electrons with momentum between $\mathbf{p}_{\alpha}$ and $\mathbf{p}_{\alpha}+d\mathbf{p}_{\alpha}$. 

For plasma with a spherically symmetric distribution function and
infinitely massive ions ($\mathbf{v}_{\beta}=0$) we can take advantage
of condition (\ref{eq:cond_1}) and write

\begin{equation}
\frac{d\mathbf{p}_{V}^{\alpha}}{dt}=\int\hbar\left(\mathbf{k}-\mathbf{q}\right)\frac{\hbar\omega_{\mathbf{k}}}{v_{\alpha}}\frac{\partial f_{\mathbf{p}_{\alpha}}^{\alpha}}{\partial p_{\alpha}}w_{\mathbf{p}_{\alpha},\mathbf{p}_{\beta}}^{Br}f_{\mathbf{p}_{\beta}}^{\beta}n_{\mathbf{k}}d\mathbf{k}d\mathbf{q}d\mathbf{p}_{\beta}d\mathbf{p}_{\alpha}.\label{eq:dpdt_recoil}
\end{equation}

\noindent This suggests that the probability of the total absorption
(inverse Bremsstrahlung plus stimulated Bremsstrahlung emission) in
plasma with a spherically symmetric distribution function is proportional
to the probability of spontaneous Bremsstrahlung emission and is $\left(\hbar\omega_{\mathbf{k}}/v_{\alpha}\right)\left(\partial\ln f_{\mathbf{p}_{\alpha}}^{\alpha}/\partial p_{\alpha}\right)w_{\mathbf{p}_{\alpha},\mathbf{p}_{\beta}}^{Br}\left(\mathbf{k},\mathbf{q}\right)$.
For plasma near equilibrium with Maxwell distribution function, which
for convenience we will consider, this probability becomes $\left(\hbar\omega_{\mathbf{k}}/T\right)w_{\mathbf{p}_{\alpha},\mathbf{p}_{\beta}}^{Br}\left(\mathbf{k},\mathbf{q}\right)$
and is actually correct even for the finite ion mass.

Consider the incoming electromagnetic radiation that consists of photons
with $\mathbf{k}=k\mathbf{e}_{z}$ and of the total intensity $I=c\int\hbar\omega n_{\mathbf{k}}d\mathbf{k}$.
Because of the condition (\ref{eq:cond_1}) the recoil momentum can
be divided into the parts parallel and perpendicular to the velocity
component: 
\begin{equation}
\mathbf{q}=-\frac{\omega-\mathbf{k}\mathbf{v}}{v^{2}}\mathbf{v}+\mathbf{q}_{\perp}.
\end{equation}

\noindent Then the rate of momentum absorption directed along the
$z$-axis can be written as

\begin{multline}
\frac{d\mathbf{p}_{V,z}^{\alpha}}{dt}=\int\hbar\left(k+\frac{\omega}{v}\frac{v_{z}}{v}-\frac{\mathbf{k}\mathbf{v}}{v}\frac{v_{z}}{v}-q_{\perp_{z}}\right)\\
\times\frac{\hbar\omega}{T}w_{\mathbf{p}_{\alpha},\mathbf{p}_{\beta}}^{Br}f_{\mathbf{p}_{\alpha}}^{\alpha}f_{\mathbf{p}_{\beta}}^{\beta}n_{\mathbf{k}}d\mathbf{k}d\mathbf{q}d\mathbf{p}_{\beta}d\mathbf{p}_{\alpha}.\label{eq:dpdt_qdecomp}
\end{multline}

\noindent To calculate the probability of Bremsstrahlung (\ref{eq:prob})
we express

\begin{multline}
\left|\left[\mathbf{e}_{k}\times\mathbf{q}\right]+\frac{\mathbf{k}\mathbf{q}}{\omega-\mathbf{k}\mathbf{v}}\left[\mathbf{e}_{k}\times\mathbf{v}\right]\right|^{2}\\
=\left|\left[\mathbf{e}_{z}\times\mathbf{q}_{\perp}\right]+\left(-\frac{\omega}{v^{2}}+\frac{\mathbf{k}\mathbf{q}_{\perp}}{\omega-\mathbf{k}\mathbf{v}}\right)\left[\mathbf{e}_{z}\times\mathbf{v}\right]\right|^{2}\\
=q_{\perp}^{2}-q_{\perp_{z}}^{2}+\frac{\omega^{2}}{v^{2}}\frac{v_{\perp}^{2}}{v^{2}}+2\frac{\omega}{v^{2}}q_{\perp_{z}}\left(v_{z}-v_{\perp}\beta_{\perp}\right)-2q_{\perp_{z}}^{2}\beta_{z},
\end{multline}

\noindent where we introduced $\boldsymbol{\beta}=\mathbf{v}/c$,
used the expression for the scalar quadruple product $\left[\mathbf{e}_{z}\times\mathbf{q}_{\perp}\right]\cdot\left[\mathbf{e}_{z}\times\mathbf{v}\right]=-q_{\perp_{z}}v_{z}$,
and kept only the first order terms.

We can write the $z$-axis projection of the perpendicular to the
velocity component of the recoil momentum as $q_{\perp_{z}}=q_{\perp}\sin\theta\sin\varphi_{q_{\perp}}$,
where $\theta$ is the angle between velocity and the $z$-axis, i.e.
$v_{z}=v\cos\theta$ and $v_{\perp}=v\sin\theta$, while $\varphi_{q_{\perp}}$
is the polar angle of $q_{\perp}$ in the plane perpendicular to $\mathbf{v}$.
We then integrate over $\varphi_{q_{\perp}}$ from $0$ to $2\pi$
and over $dq_{\parallel}q_{\perp}dq_{\perp}$. When we integrate over
$dq_{\perp}$ it is necessary to introduce a cutoff to get rid of
a logarithmic divergence. For definiteness, we will use the quantum
mechanical cutoff ($q_{max}=m_{\alpha}v/\hbar$), which is correct
when the Born approximation can be applied ($v\gg e^{2}/\hbar$).
In the opposite classical limit ($v\ll e^{2}/\hbar$) the proper cutoff
is $q_{max}=m_{\alpha}v^{2}/e_{\alpha}e_{\beta}$ and the conclusions
of the paper should remain true but all logarithmic factors should
be replaced with $\ln\left(m_{\alpha}v^{3}/\omega e_{\alpha}e_{\beta}\right)$.

Keeping only the leading logarithmic terms, the probability of Bremsstrahlung
integrated over $d\mathbf{q}$ is then

\begin{multline}
\int w_{\mathbf{p}_{\alpha},\mathbf{p}_{\beta}}^{Br}\left(\mathbf{k},\mathbf{q}\right)d\mathbf{q}\\
\approx\frac{e_{\alpha}^{4}e_{\beta}^{2}}{\pi\hbar m_{\alpha}^{2}\omega^{3}v}\left(1+\frac{v_{z}^{2}}{v^{2}}+4\beta_{z}\frac{v_{z}^{2}}{v^{2}}\right)\ln\left(\frac{m_{\alpha}v^{2}}{\hbar\omega}\right),\label{eq:prob_integrated}
\end{multline}

\noindent which determines the absorbed power, and

\begin{multline}
\int q_{\perp_{z}}w_{\mathbf{p}_{\alpha},\mathbf{p}_{\beta}}^{Br}\left(\mathbf{k},\mathbf{q}\right)d\mathbf{q}\\
\approx\frac{\omega}{c}\frac{e_{\alpha}^{4}e_{\beta}^{2}}{\pi\hbar m_{\alpha}^{2}\omega^{3}v}2\frac{v_{\perp}^{2}}{v^{2}}\left(\frac{cv_{z}}{v^{2}}+2\frac{v_{z}^{2}}{v^{2}}-\frac{v_{\perp}^{2}}{v^{2}}\right)\ln\left(\frac{m_{\alpha}v^{2}}{\hbar\omega}\right),\label{eq:q_perb_prob_integrated}
\end{multline}

\noindent which determines the amount of momentum change in the direction
perpendicular to the velocity. This is needed to calculate the current.
Note that while it is not necessary to retain the first order terms
in Eq.~(\ref{eq:prob_integrated}) to calculate the absorbed power,
one needs to keep them while calculating current. Note also in Eq.~(\ref{eq:prob_integrated})
that electrons moving in the direction of the photon $(\beta_{z}>0)$
are more likely to absrob energy than electrons moving in the opposite
direction $(\beta_{z}<0)$. This is consistent with the picture that
an electron moving in the direction of the photon can absorb its energy
through a smaller angle scatter than would an electron moving in the
opposite direction.

From Eqs.~(\ref{eq:prob_integrated}) and~(\ref{eq:q_perb_prob_integrated})
we can write the rate of momentum absorption as:

\begin{multline}
\frac{d\mathbf{p}_{V,z}^{\alpha}}{dt}=\int\frac{\hbar\omega}{c}\left(1+\frac{cv_{z}}{v^{2}}-\frac{v_{z}^{2}}{v^{2}}\right)\frac{\hbar\omega}{T}\\
\times\frac{n_{\beta}e_{\alpha}^{4}e_{\beta}^{2}}{\pi\hbar m_{\alpha}^{2}\omega^{3}v}\left(1+\frac{v_{z}^{2}}{v^{2}}+4\beta_{z}\frac{v_{z}^{2}}{v^{2}}\right)\ln\left(\frac{m_{\alpha}v^{2}}{\hbar\omega}\right)f_{\mathbf{p}}^{\alpha}d\mathbf{p}n_{\mathbf{k}}d\mathbf{k}\\
-\int\frac{\hbar\omega}{c}\frac{\hbar\omega}{T}\frac{n_{\beta}e_{\alpha}^{4}e_{\beta}^{2}}{\pi\hbar m_{\alpha}^{2}\omega^{3}v}\\
\times2\frac{v_{\perp}^{2}}{v^{2}}\left(\frac{cv_{z}}{v^{2}}+2\frac{v_{z}^{2}}{v^{2}}-\frac{v_{\perp}^{2}}{v^{2}}\right)\ln\left(\frac{m_{\alpha}v^{2}}{\hbar\omega}\right)f_{\mathbf{p}}^{\alpha}d\mathbf{p}n_{\mathbf{k}}d\mathbf{k}.
\end{multline}

Integrating over angle $\theta$ we get

\begin{equation}
\frac{d\mathbf{p}_{V,z}^{\alpha}}{dt}=\frac{32}{15}\int\frac{\hbar\omega}{T}\frac{n_{\beta}e_{\alpha}^{4}e_{\beta}^{2}}{\pi cm_{\alpha}^{2}\omega^{2}v}\ln\left(\frac{m_{\alpha}v^{2}}{\hbar\omega}\right)f_{v}^{\alpha}d\mathbf{v}n_{\mathbf{k}}d\mathbf{k}.
\end{equation}

\noindent Therefore,

\begin{equation}
\frac{d\mathbf{p}_{V,z}^{\alpha}}{dt}=\frac{8}{5}\frac{\alpha I}{c}.\label{eq:dpdt}
\end{equation}

\noindent Here $\alpha$ is the effective absorption coefficient:

\begin{equation}
\alpha\approx\frac{4}{3}\sqrt{\frac{2}{\pi}}\frac{n_{\alpha}n_{\beta}e_{\alpha}^{4}e_{\beta}^{2}}{\pi cm_{\alpha}^{3}\omega^{2}v_{th}^{3}}\ln\left(\frac{2T}{\hbar\omega}\right),
\end{equation}

\noindent where $v_{th}^{2}=T/m_{\alpha}$. This absorption coefficient
determines the total absorbed power density: $P_{V}^{abs}=\alpha I$.

If we ignored the recoil momentum and assumed that electrons absorb
just the incoming photon momentum $\hbar\mathbf{k}$, then the rate
of momentum change would be:

\begin{equation}
\frac{d\mathbf{p}_{V,z}^{\mathbf{k}}}{dt}=\int\hbar k\frac{\hbar\omega}{T}w_{\mathbf{p}_{\alpha},\mathbf{p}_{\beta}}^{\alpha,\beta}f_{\mathbf{p}_{\alpha}}^{\alpha}f_{\mathbf{p}_{\beta}}^{\beta}n_{\mathbf{k}}d\mathbf{k}d\mathbf{q}d\mathbf{p}_{\beta}d\mathbf{p}_{\alpha}=\frac{\alpha I}{c}.\label{eq:dpdt_naive}
\end{equation}

Thus, due to the recoil, electrons get $8/5$ times more momentum
than they would have got absorbing only the photon momentum, which
is consistent with the result obtained in \cite{Pashinin1978}. This
conclusion is true for any spherically symmetric distribution function,
not just a Maxwellian. This additional momentum absorbed by electrons
(as a whole) is in the direction of the incoming radiation. The ions
(as a whole), on the other hand, absorb momentum in the opposite to
the incoming radiation direction such that the total rate of momentum
absorption for plasma is equal to the rate of photon momentum absorption:

\begin{equation}
\frac{d\mathbf{p}_{V,z}^{\alpha}}{dt}+\frac{d\mathbf{p}_{V,z}^{\beta}}{dt}=\frac{8}{5}\frac{\alpha I}{c}-\frac{3}{5}\frac{\alpha I}{c}=\frac{d\mathbf{p}_{V,z}^{\mathbf{k}}}{dt}=\hbar k\frac{dN_{ph}^{abs}}{dt}=\frac{\alpha I}{c}.
\end{equation}

It is curious that after averaging for spherically symmetric distribution
functions the last two terms in Eq. (\ref{eq:dpdt_qdecomp}) cancel
each other and the rate of momentum absorption becomes just

\begin{multline}
\frac{d\mathbf{p}_{V,z}^{\alpha}}{dt}=\int\hbar\left(k+\frac{\omega}{v}\frac{v_{z}}{v}\right)\\
\times\frac{\hbar\omega}{T}w_{\mathbf{p}_{\alpha},\mathbf{p}_{\beta}}^{Br}f_{\mathbf{p}_{\alpha}}^{\alpha}f_{\mathbf{p}_{\beta}}^{\beta}n_{\mathbf{k}}d\mathbf{k}d\mathbf{q}d\mathbf{p}_{\beta}d\mathbf{p}_{\alpha},\label{eq:dpdt_sph}
\end{multline}

\noindent where integration of $w_{\mathbf{p}_{\alpha},\mathbf{p}_{\beta}}^{Br}\left(\mathbf{k},\mathbf{q}\right)$
over $d\mathbf{q}$ can be done independently to get~(\ref{eq:prob_integrated}).
$\int w_{\mathbf{p}_{\alpha},\mathbf{p}_{\beta}}^{Br}\left(\mathbf{k},\mathbf{q}\right)d\mathbf{q}$
has a zero order term, which is even in $v_{z}$, and a first order
term $O\left(\beta_{z}\right)$, which is odd in $v_{z}$. In Eq.~(\ref{eq:dpdt_sph})
the first term $k=\omega/c$ is the momentum of the absorbed photon
and it is much smaller than the momentum coming from the recoil $\left(\omega/v\right)\left(v_{z}/v\right)$.
However, the photon term $k=\omega/c$ is the same for all electrons
and is multiplied by the zero order term in $\int w_{\mathbf{p}_{\alpha},\mathbf{p}_{\beta}}^{Br}\left(\mathbf{k},\mathbf{q}\right)d\mathbf{q}$,
while the recoil term, which depends on the velocity projection $v_{z}$,
has contribution only from the first order term in $\int w_{\mathbf{p}_{\alpha},\mathbf{p}_{\beta}}^{Br}\left(\mathbf{k},\mathbf{q}\right)d\mathbf{q}$,
because the zero order term is the same for oppositely going electrons
and so gives zero contribution after averaging over the distribution
function. Thus, after multiplication by the probability both terms
give contributions of equal order. The coefficient next to the first
order term in $\int w_{\mathbf{p}_{\alpha},\mathbf{p}_{\beta}}^{Br}\left(\mathbf{k},\mathbf{q}\right)d\mathbf{q}$
is positive, which comes from the fact that Bremsstrahlung emission
is the most pronounced in the direction of the electron velocity \cite{Jackson1999}.
Since also the recoil term is proportional to $v_{z}$, we can immediately
conclude that the averaged momentum gained by electrons due to the
recoil is in the positive $z$-axis direction.

\section{Inverse Bremsstrahlung Current}

The time evolution of the current density can be put as

\begin{equation}
\frac{d\mathbf{j}}{dt}=-\frac{e}{m_{e}}\frac{d\mathbf{p}_{V}^{e}}{dt}-\nu_{Sp}\mathbf{j}.\label{eq:current_evol_naive}
\end{equation}

\noindent This is a fluid approach, since it takes into account only
how much momentum is absorbed by electrons, not which electrons absorb
the momentum.

The collision frequency $\nu_{Sp}$ in Eq.~(\ref{eq:current_evol_naive})
corresponds to the Spitzer conductivity and can be approximated by
the following empirical formula \cite{Diver2001}:

\begin{equation}
\nu_{Sp}=\frac{Z}{3}\sqrt{\frac{2}{\pi}}\left(0.295+\frac{0.39}{0.85+Z}\right)\frac{\Gamma}{v_{th}^{3}},
\end{equation}

\noindent where $\Gamma=\omega_{p}^{4}\ln\Lambda/4\pi n$ and $Z$
is the ion charge. From Eq.~(\ref{eq:current_evol_naive}) the stationary
current density is

\begin{equation}
\mathbf{j}_{fluid}=-\frac{e}{m_{e}}\nu_{Sp}^{-1}\frac{d\mathbf{p}_{V}^{e}}{dt}.\label{eq:j}
\end{equation}

Since the current density in the fluid approximation is proportional
to the rate of momentum absorption, the current corrected for the
recoil is $8/5$ times higher than the simple fluid estimate ignoring
the recoil, and is equal to

\noindent 
\begin{equation}
j_{fluid}=-\frac{8}{5}\frac{e}{m_{e}}\frac{\alpha I}{c}\nu_{Sp}^{-1}=-\frac{20.4}{Z\left(1+\frac{1.32}{0.85+Z}\right)}\frac{ev_{th}^{3}}{m_{e}\Gamma}\frac{\alpha I}{c}.\label{eq:j_fluid}
\end{equation}

\begin{figure}
\includegraphics[width=0.45\textwidth]{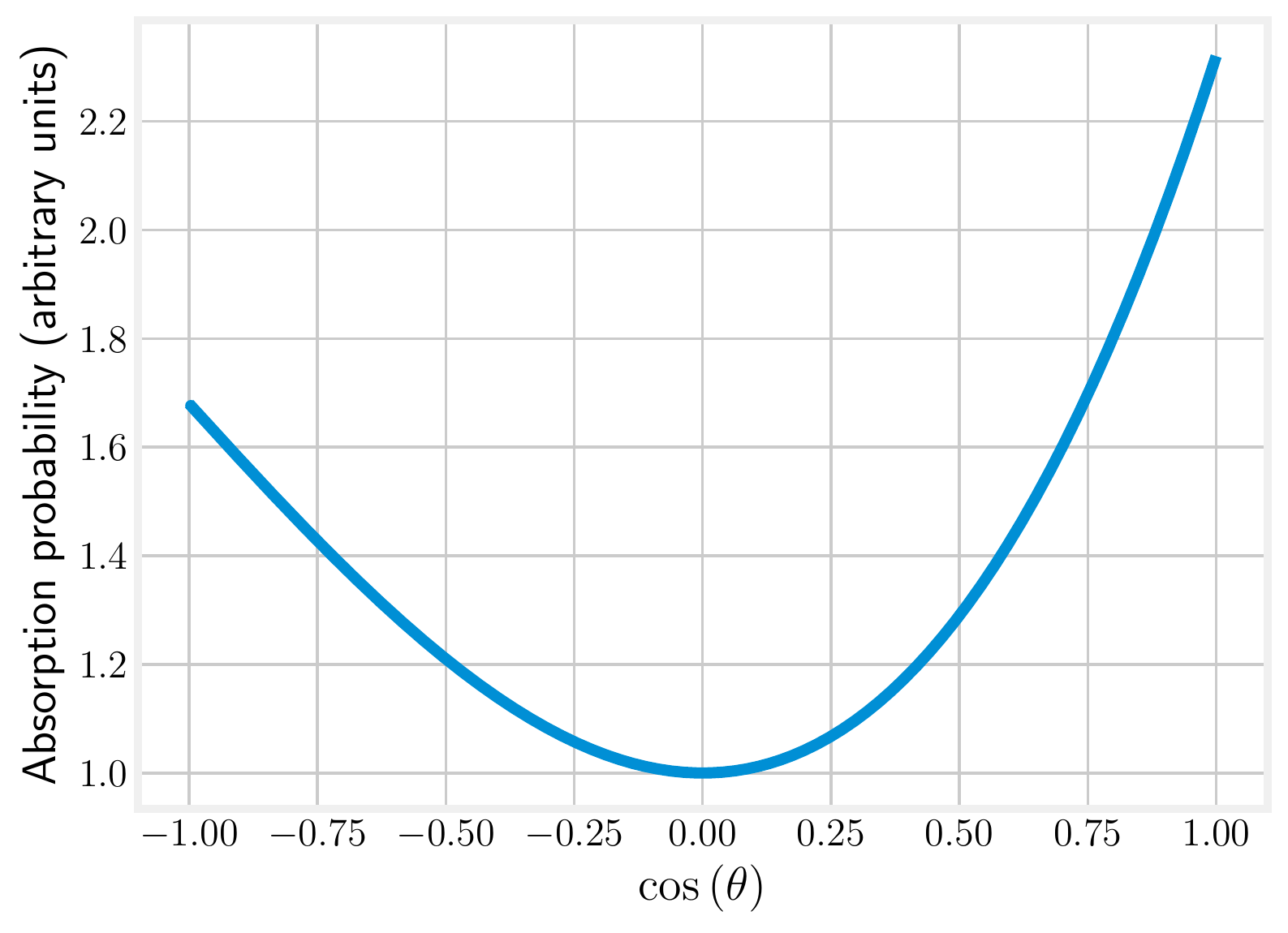}

\caption{\label{fig02}The probability of Bremsstrahlung absorption in arbitrary
units versus the angle between the electron velocity and the incoming
photon direction $\cos\theta=v_{z}/v$ for $\beta=0.08$.}
\end{figure}

However, the Spitzer conductivity is strictly applicable only to the
current produced by dc electric field, when all electrons get equal
acceleration in the same direction. The current generation due to
inverse Bremsstrahlung is not equivalent to the action of dc electric
field because different electrons absorb different amount of power
and are pushed in different directions. 

One example of the kinetic effects is the additional current due to
asymmetric absorption of radiation. Fig.~\ref{fig02} shows the integrated
probability of absorption within $d\theta$ given by Eq.~(\ref{eq:prob_integrated})
for electrons lying on the circle with radius $\beta=0.08$ in velocity
space. We see that the electrons going in the direction of the incoming
photons ($0\leq\theta<\pi/2$) absorb more radiation than electrons
going in the opposite direction ($\pi/2<\theta\leq\pi$). This asymmetric
absorption will create additional current because the collision frequency
in plasma is speed dependent and thus electrons going in the direction
of the incoming radiation will experience less resistance from the
plasma than electrons going in the opposite direction resulting in
more current.

\begin{figure}
\includegraphics[width=1\columnwidth]{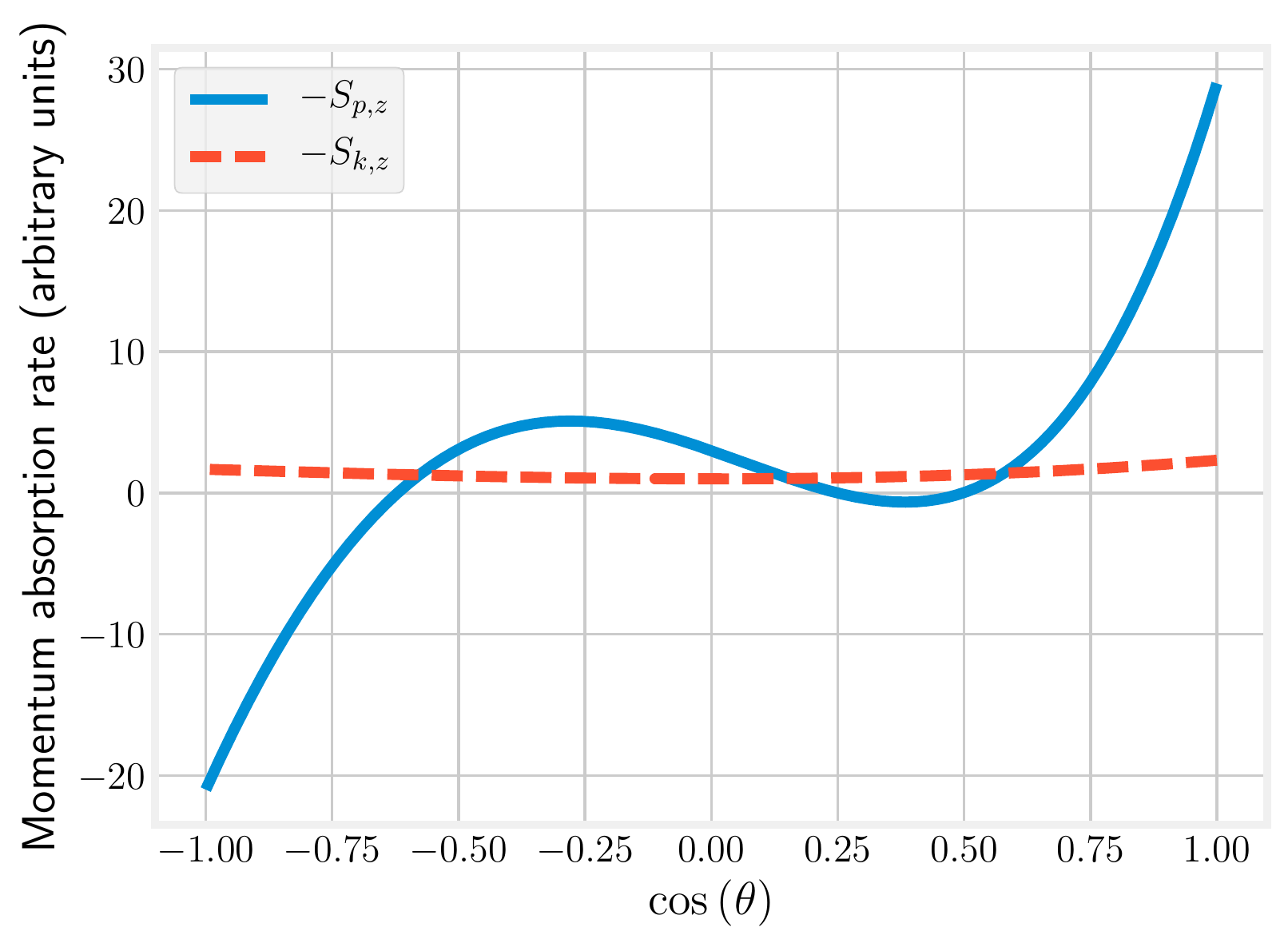}

\caption{\label{fig03}The momentum absorption rate per electron as a function
of $\cos\theta=v_{z}/v$ for $\beta=0.08$: along the $z$-axis taking
into account the recoil (solid blue), along the $z$-axis taking into
account only the photon momentum (dashed red).}
\end{figure}

Fig.~\ref{fig03} shows, averaged over all possible recoils, the
rate of momentum absorption along the $z$-axis by an electron with
$\beta=0.08$ versus $\cos\theta=v_{z}/v$. $-S_{p,z}$ is defined
by Eq.~(\ref{eq:dpdt_recoil}) and determines the rate of momentum
absorption taking into account the recoil effect. $-S_{k,z}$ is defined
by Eq.~(\ref{eq:dpdt_naive}) and determines the rate of momentum
absorption assuming that only the photon momentum is absorbed. We
can see that the recoil effect not only changes the integrated (average)
rate of momentum absorption but radically alters the distribution
of the absorbed momentum in velocity space. For $-S_{k,z}$ the momentum
absorption rate is always positive, i.e. along the $z$-axis, and
does not strongly depend on $\cos\theta$, while for $-S_{p,z}$ the
momentum absorption rate varies greatly with $\cos\theta$ both in
magnitude and sign. In considering Bremsstrahlung absorption by a
particular electron, the natural directions are along the electron
velocity and perpendicular to the electron velocity. When $\left|\cos\left(\theta\right)\right|$
is close to 1, the velocity of the electron is either parallel or
antiparallel to the direction of the incoming photon and so the change
in momentum along the $z$-axis is determined mostly by the recoil
parallel to the velocity, which is about $\left(\hbar\omega/v\right)\left(v_{z}/v\right)$
in each act of the Bremsstrahlung, as was shown previously. For smaller
values of $\left|\cos\left(\theta\right)\right|$ the change in momentum
along the $z$-axis is mostly determined by the recoil perpendicular
to the electron velocity. This is why the absorption rate shown in
Fig.~\ref{fig03} changes sign.

In general, the distribution function will evolve both under the influence
of Bremsstrahlung absorption and under the influence of collisions:

\begin{equation}
\frac{\partial f_{\mathbf{p}}^{e}}{\partial t}=\left(\frac{\partial f_{\mathbf{p}}^{e}}{\partial t}\right)_{Br}+\left(\frac{\partial f_{\mathbf{p}}^{e}}{\partial t}\right)_{coll},
\end{equation}

\noindent and the time-evolution of the current should be described
more completely than Eq.~(\ref{eq:current_evol_naive}) does by
\begin{equation}
\frac{d\mathbf{j}}{dt}=-e\int\mathbf{v}\frac{\partial f_{\mathbf{p}}^{e}}{\partial t}d\mathbf{p}.
\end{equation}

Following \cite{Fisch1987} we can write the current density at time
$t$ as the rate of pushing electrons times the ensemble-averaged
current difference:

\begin{multline}
j_{cd}\left(t\right)=\sum_{\mathbf{v},\triangle\mathbf{v}}\int_{0}^{t}d\tau\frac{P_{V}\left(\tau,\mathbf{v},\triangle\mathbf{v}\right)}{\triangle\varepsilon}\\
\times\left\langle q_{e}v_{z}\left(t-\tau,\mathbf{v}+\triangle\mathbf{v}\right)-q_{e}v_{z}\left(t-\tau,\mathbf{v}\right)\right\rangle \\
\underset{\triangle\mathbf{v}\rightarrow0}{=}\sum_{\mathbf{v},\triangle\mathbf{v}}\int_{0}^{t}d\tau\frac{P_{V}\left(\tau,\mathbf{v},\triangle\mathbf{v}\right)}{\triangle\varepsilon}\triangle\mathbf{v}\cdot\frac{\partial\left\langle qv_{z}\left(t-\tau,\mathbf{v}\right)\right\rangle }{\partial\mathbf{v}}\label{eq:j_t_cd}
\end{multline}

If the power is independent of time we can put integration inside
the ensemble-averaged current and write for a steady-state current:

\begin{equation}
j_{cd}=\int\left[-\frac{e}{m_{e}}\frac{\hbar\left(\mathbf{k}-\mathbf{q}\right)\cdot\partial\chi/\partial\mathbf{v}}{\hbar\omega}\right]dP_{V}\left(\mathbf{v},\mathbf{k},\mathbf{q}\right),\label{eq:j_cd_int}
\end{equation}
where we expressed infinitesimal changes in energy and velocity through
$\omega,$ $\mathbf{k}$, $\mathbf{q}$, changed from summation to
integration, and introduced a Green's function: $\chi=\int_{0}^{\infty}\left\langle v_{z}\left(\tau,\mathbf{v}\right)\right\rangle d\tau$.
In most cases it is possible to express the Green's function as $\chi(\mathbf{v})=v_{z}\nu^{-1}\left(v\right)$,
where $\nu^{-1}$ can be thought of as an effective collision frequency
\cite{FidoneGranataJohner1988}.

The expression in square brackets of Eq.~(\ref{eq:j_cd_int}) can
be understood as incremental current drive efficiency. Thus, to find
the generated current one needs to average the incremental current
drive efficiency over the power density absorbed:

\begin{multline}
j_{cd}=\int\left(\frac{\delta j_{z}}{\delta P_{V}}\right)dP_{V}\\
=\frac{e}{m_{e}}\int\frac{\left(\mathbf{k}-\mathbf{q}\right)\cdot\partial\chi/\partial\mathbf{v}}{\omega}\\
\times\frac{m_{e}v^{2}}{2}\frac{\partial}{\partial\mathbf{v}}\cdot\hbar\left(\mathbf{k}-\mathbf{q}\right)\frac{\hbar\omega}{m_{e}T}w_{\mathbf{p}_{\alpha},\mathbf{p}_{\beta}}^{Br}f_{\mathbf{p}_{\alpha}}^{\alpha}f_{\mathbf{p}_{\beta}}^{\beta}n_{\mathbf{k}}d\mathbf{k}d\mathbf{q}d\mathbf{p}_{\beta}d\mathbf{p}_{\alpha}\\
=\frac{e}{m_{e}}\int\left[\frac{\nu^{-1}}{\omega}\left(k_{z}-q_{z}\right)+\frac{\partial\nu^{-1}}{\partial v}\frac{v_{z}}{v}\right]\\
\times\frac{m_{e}v^{2}}{2}\frac{\partial}{\partial\mathbf{v}}\cdot\hbar\left(\mathbf{k}-\mathbf{q}\right)\frac{\hbar\omega}{m_{e}T}w_{\mathbf{p}_{\alpha},\mathbf{p}_{\beta}}^{Br}f_{\mathbf{p}_{\alpha}}^{\alpha}f_{\mathbf{p}_{\beta}}^{\beta}n_{\mathbf{k}}d\mathbf{k}d\mathbf{q}d\mathbf{p}_{\beta}d\mathbf{p}_{\alpha}.\label{eq:j_cd}
\end{multline}

The first term in square brackets of Eq.~(\ref{eq:j_cd}), which
is proportional to $k_{z}-q_{z}$, is the usual current due to momentum
injection along the $z$-axis, while the second term, which is proportional
to $\partial\nu^{-1}/\partial v$, is the current due to asymmetric
absorption.

One might want to calculate the generated current by summing the incremental
currents instead:

\begin{multline}
j_{cd,res}=\int\delta j_{z}=-\frac{e}{m}\int\hbar\left(\mathbf{k}-\mathbf{q}\right)\cdot\frac{\partial\chi}{\partial\mathbf{v}}\\
\times\frac{\hbar\omega}{T}w_{\mathbf{p}_{\alpha},\mathbf{p}_{\beta}}^{Br}f_{\mathbf{p}_{\alpha}}^{\alpha}f_{\mathbf{p}_{\beta}}^{\beta}n_{\mathbf{k}}d\mathbf{k}d\mathbf{q}d\mathbf{p}_{\beta}d\mathbf{p}_{\alpha}\\
=e\int\mathbf{S}_{\mathbf{v}}\cdot\frac{\partial\chi}{\partial\mathbf{v}}d\mathbf{v},\label{eq:j_cd_naive}
\end{multline}

\noindent where we used the wave induced flux in velocity space $\mathbf{S_{\mathbf{v}}}=m_{e}^{2}\mathbf{S}_{\mathbf{p}}$.
Eq.~(\ref{eq:j_cd_naive}) follows from Eq.~(\ref{eq:j_t_cd}) if
the power absorbed is localized around certain velocity. Therefore,
Eqs.~(\ref{eq:j_cd}) and~(\ref{eq:j_cd_naive}) are identical when
the absorption is localized in the velocity space, but they produce
different results otherwise. In the present problem all electrons
are pushed by the incoming electromagnetic field and Eq.~(\ref{eq:j_cd_naive})
miscalculates the generated current density.

After integration by parts, Eq.~(\ref{eq:j_cd}) can be written as

\begin{multline}
j_{cd}=-\frac{e}{2}\int\frac{\partial\left(v\frac{\partial\nu^{-1}}{\partial v}\right)}{\partial v}\frac{v_{z}}{v}\hbar\omega\frac{\hbar\omega}{m_{e}T}N_{ph}n_{\beta}w_{\mathbf{p}_{\alpha},\mathbf{p}_{\beta}}^{Br}f_{\mathbf{v}}^{e}d\mathbf{q}d\mathbf{v}\\
-e\int\frac{\partial\left(v\nu^{-1}\right)}{\partial v}\hbar\left(k_{z}-q_{z}\right)\frac{\hbar\omega}{m_{e}T}N_{ph}n_{\beta}w_{\mathbf{p}_{\alpha},\mathbf{p}_{\beta}}^{Br}f_{\mathbf{v}}^{e}d\mathbf{q}d\mathbf{v}.
\end{multline}

The Green's function and the corresponding effective collision frequency
$\nu$, generally speaking, can be found only numerically. However,
the high-velocity approximation exists \cite{KarneyFisch1985,Fisch1987}:

\begin{equation}
\nu^{-1}=\frac{v^{3}}{\Gamma\left(5+Z\right)}+\frac{9v_{th}^{2}v}{\Gamma\left(5+Z\right)\left(3+Z\right)}.\label{eq:freq_inv}
\end{equation}

This expression has two shortcomings. First, it uses the high-velocity
approximation both for electron-electron and electron-ion collisions.
While for electron-ion collisions this approximation is always good,
it is less so for electron-electron collisions. Since it is mostly
thermal electrons that absorb through Bremsstrahlung, the high-velocity
approximation will noticeably underestimate the current for low $Z$
plasma. Second, this expression violates the momentum conservation
in electron-electron collisions. Thus, we expect that Eq.~(\ref{eq:freq_inv})
is a good approximation for high $Z$ plasma, but for low $Z$ plasma
the error in the current can be appreciable.

After straightforward calculations using $\nu$ defined by Eq.~(\ref{eq:freq_inv})
we obtain from Eq.~(\ref{eq:j_cd}):

\begin{equation}
j_{cd}=-\frac{34.2}{5+Z}\frac{ev_{th}^{3}}{m_{e}\Gamma}\frac{\alpha I}{c}-\frac{39.5}{\left(5+Z\right)\left(3+Z\right)}\frac{ev_{th}^{3}}{m_{e}\Gamma}\frac{\alpha I}{c},\label{eq:j_cd_fin}
\end{equation}

\noindent while Eq.~(\ref{eq:j_cd_naive}) would only give factors
12.8 and 24.8 respectively in the above formula.

For comparison, in the fluid approximation the current density corrected
for the recoil, which is given by Eq.~(\ref{eq:j_fluid}), can be
represented as

\begin{equation}
j_{fluid}=e\nu_{Sp}^{-1}\int S_{v,z}d\mathbf{v}.
\end{equation}

\noindent We can clearly see that Eq.~(\ref{eq:j_cd}) has an additional
term that is responsible for the current due to asymmetric absorption.

Because of the use of the high-velocity and momentum conservation
violating approximation for $\nu$, Eq.~(\ref{eq:j_cd_fin}) underestimates
the current, especially for small $Z$. Reckoning that electron-electron
collisions conserve current, to remedy this problem we propose an
alternative hybrid expression, where the part of the current in Eq.~(\ref{eq:j_cd})
proportional to $k_{z}-q_{z}$ is substituted by the fluid expression
Eq.~(\ref{eq:j_fluid}), while the part proportional to $\partial\nu^{-1}/\partial v$
is left unchanged:

\begin{multline}
j_{hybrid}=j_{fluid}-\frac{e}{m_{e}}\int\frac{\partial\nu^{-1}}{\partial v}\frac{v_{z}}{v}dP_{V}\left(\mathbf{v},\mathbf{k},\mathbf{q}\right)\\
=j_{fluid}-\frac{19.2}{5+Z}\frac{ev_{th}^{3}}{m_{e}\Gamma}\frac{\alpha I}{c}-\frac{12.4}{\left(5+Z\right)\left(3+Z\right)}\frac{ev_{th}^{3}}{m_{e}\Gamma}\frac{\alpha I}{c}.\label{eq:j_hybrid}
\end{multline}

If all electrons were to absorb equal amount of power, then the part
of the current in Eq.~(\ref{eq:j_cd}) proportional to $k_{z}-q_{z}$
would be exactly given by the fluid expression Eq.~(\ref{eq:j_fluid}).
In case of Bremsstrahlung absorption it is mostly thermal electrons
that absorb radiation and the fluid formula overestimates the corresponding
part of the current. On the other hand, the second part of Eq.~(\ref{eq:j_hybrid})
underestimates the current because of the high-velocity limit for
$\nu$. So all in all, Eq.~(\ref{eq:j_hybrid}) can be a decent approximation
for the current for all values of $Z$.

\begin{figure}
\includegraphics[width=1\columnwidth]{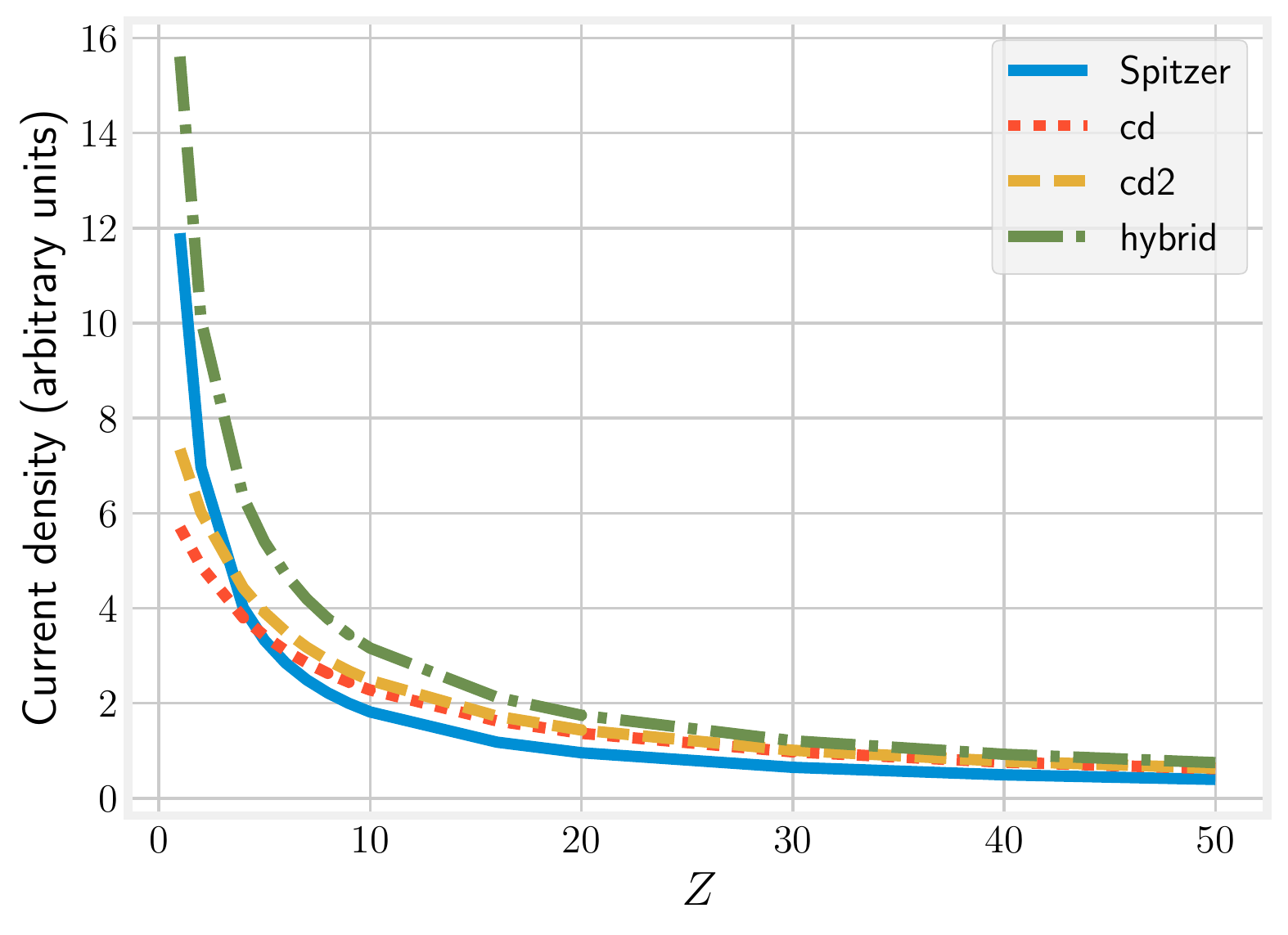}

\caption{\label{fig04}The generated current density versus the ion charge
$Z$: fluid approximation with the Spitzer conductivity given by Eq.~(\ref{eq:j_fluid})
(solid blue), current drive approximation keeping only the first term
in Eq.~(\ref{eq:j_cd_fin}) (dotted red), current drive approximation
keeping both terms in Eq.~(\ref{eq:j_cd_fin}) (dashed orange), hybrid
current given by Eq.~(\ref{eq:j_hybrid}) (dash-dotted green).}
\end{figure}

Fig.~\ref{fig04} shows the generated current given by the fluid
formula~(\ref{eq:j_fluid}), by the current drive formula~(\ref{eq:j_cd_fin})
keeping one and two terms in Eq.~(\ref{eq:j_cd_fin}), and by the
hybrid expression~(\ref{eq:j_hybrid}) versus the ion charge $Z$.
We see that for small $Z$ the current drive formula substantially
underestimates current making it even lower than the fluid prediction.
However, starting already with $Z=4$ the current drive estimate~(\ref{eq:j_cd_fin})
gives higher current. For higher $Z$, when electron-electron collisions
become negligible, the ratio of the current drive prediction to the
Spitzer becomes stable and for infinite $Z$ is around 1.7, so that
for high $Z$ the generated current with the recoil and kinetic effects
taken into account is at least 2.7 higher than the naive fluid estimate
without recoil would suggest. The hybrid expression is 1.3 times larger
than the fluid estimate even for $Z=1$ and for $Z$ going to infinity
the increase is about 2. To get better and definite results for small
$Z$ plasma it is necessary to use more accurate than Eq.~(\ref{eq:freq_inv})
estimate of the effective collision frequency $\nu$ or perform computer
simulations.

\section{Summary}

We analytically considered the generation of the plasma current resulting
from electron-ion Bremsstrahlung absorption using the following approximations:
the polarization effects in Bremsstrahlung are negligible; velocities
are non-relativistic; recoil and photon momenta are small in comparison
with the electron momentum; ions have infinite mass; waves are electromagnetic
with the dispersion relation $\omega=kc$; and the plasma dielectric
function is close to one. The laser intensity is not too high, so
that the quiver velocity $eE/m\omega$ is much smaller than the thermal
velocity. We also note that the logarithmic dependence on velocity
has been ignored throughout the paper and $\ln\left(m_{\alpha}v^{2}/\hbar\omega\right)$
has been substituted by $\ln\left(2T/\hbar\omega\right)$ in all the
equations.

We investigated how the momentum and energy are absorbed by electrons
within the velocity space and confirmed the result obtained in \cite{Pashinin1978},
namely that the averaged momentum absorption by electrons with the
recoil taken into account is $8/5$ times higher than the momentum
absorption assuming that electrons absorb just the photon momentum.
In addition, we demonstrated that for high $Z$ plasma the actual
current with the kinetic effects taken into account is at least 2.7
times higher than the naive fluid estimates without recoil would suggest,
both because electrons get the recoil momentum from the Coulomb field
of ions during the absorption and because electrons absorb power asymmetrically.
We also proposed a hybrid expression of fluid and kinetic descriptions
for the current that can be a good approximation for all values of
$Z$.

The calculation of the current generated from Bremsstrahlung absorption
is a fundamental problem of the basic plasma physics. Thus, the results
here ought to be of interest in the different areas where radiation
driven currents and the generated magnetic fields are important. Areas
in which these effects might be important include the radiation driven
magnetic field in astrophysics \cite{Munirov2017a,Widrow2002,Durrive2017}
and laboratory experiments that use lasers to drive current \cite{Kruer1988},
in particular for applications to inertial confinement fusion.

\begin{acknowledgments}
This work was supported by NNSA Grant No. DENA0002948.
\end{acknowledgments}

\bibliographystyle{apsrev4-1}
\bibliography{bremsstrahlung}

\begin{thebibliography}{14}%
\makeatletter
\providecommand \@ifxundefined [1]{%
 \@ifx{#1\undefined}
}%
\providecommand \@ifnum [1]{%
 \ifnum #1\expandafter \@firstoftwo
 \else \expandafter \@secondoftwo
 \fi
}%
\providecommand \@ifx [1]{%
 \ifx #1\expandafter \@firstoftwo
 \else \expandafter \@secondoftwo
 \fi
}%
\providecommand \natexlab [1]{#1}%
\providecommand \enquote  [1]{``#1''}%
\providecommand \bibnamefont  [1]{#1}%
\providecommand \bibfnamefont [1]{#1}%
\providecommand \citenamefont [1]{#1}%
\providecommand \href@noop [0]{\@secondoftwo}%
\providecommand \href [0]{\begingroup \@sanitize@url \@href}%
\providecommand \@href[1]{\@@startlink{#1}\@@href}%
\providecommand \@@href[1]{\endgroup#1\@@endlink}%
\providecommand \@sanitize@url [0]{\catcode `\\12\catcode `\$12\catcode
  `\&12\catcode `\#12\catcode `\^12\catcode `\_12\catcode `\%12\relax}%
\providecommand \@@startlink[1]{}%
\providecommand \@@endlink[0]{}%
\providecommand \url  [0]{\begingroup\@sanitize@url \@url }%
\providecommand \@url [1]{\endgroup\@href {#1}{\urlprefix }}%
\providecommand \urlprefix  [0]{URL }%
\providecommand \Eprint [0]{\href }%
\providecommand \doibase [0]{http://dx.doi.org/}%
\providecommand \selectlanguage [0]{\@gobble}%
\providecommand \bibinfo  [0]{\@secondoftwo}%
\providecommand \bibfield  [0]{\@secondoftwo}%
\providecommand \translation [1]{[#1]}%
\providecommand \BibitemOpen [0]{}%
\providecommand \bibitemStop [0]{}%
\providecommand \bibitemNoStop [0]{.\EOS\space}%
\providecommand \EOS [0]{\spacefactor3000\relax}%
\providecommand \BibitemShut  [1]{\csname bibitem#1\endcsname}%
\let\auto@bib@innerbib\@empty
\bibitem [{\citenamefont {Pashinin}\ and\ \citenamefont
  {Fedorov}(1978)}]{Pashinin1978}%
  \BibitemOpen
  \bibfield  {author} {\bibinfo {author} {\bibfnamefont {P.~P.}\ \bibnamefont
  {Pashinin}}\ and\ \bibinfo {author} {\bibfnamefont {M.~V.}\ \bibnamefont
  {Fedorov}},\ }\href@noop {} {\bibfield  {journal} {\bibinfo  {journal} {Sov.
  J. Exp. Theor. Phys.}\ }\textbf {\bibinfo {volume} {48}},\ \bibinfo {pages}
  {228} (\bibinfo {year} {1978})}\BibitemShut {NoStop}%
\bibitem [{\citenamefont {Fisch}\ and\ \citenamefont
  {Boozer}(1980)}]{FischBoozer1980}%
  \BibitemOpen
  \bibfield  {author} {\bibinfo {author} {\bibfnamefont {N.~J.}\ \bibnamefont
  {Fisch}}\ and\ \bibinfo {author} {\bibfnamefont {A.~H.}\ \bibnamefont
  {Boozer}},\ }\href {\doibase 10.1103/PhysRevLett.45.720} {\bibfield
  {journal} {\bibinfo  {journal} {Phys. Rev. Lett.}\ }\textbf {\bibinfo
  {volume} {45}},\ \bibinfo {pages} {720} (\bibinfo {year} {1980})}\BibitemShut
  {NoStop}%
\bibitem [{\citenamefont {Fisch}(1987)}]{Fisch1987}%
  \BibitemOpen
  \bibfield  {author} {\bibinfo {author} {\bibfnamefont {N.~J.}\ \bibnamefont
  {Fisch}},\ }\href {\doibase 10.1103/RevModPhys.59.175} {\bibfield  {journal}
  {\bibinfo  {journal} {Rev. Mod. Phys.}\ }\textbf {\bibinfo {volume} {59}},\
  \bibinfo {pages} {175} (\bibinfo {year} {1987})}\BibitemShut {NoStop}%
\bibitem [{\citenamefont {Tsytovich}\ and\ \citenamefont
  {Oiringel}(1992)}]{Tsytovich1992}%
  \BibitemOpen
  \bibfield  {author} {\bibinfo {author} {\bibfnamefont {V.~N.}\ \bibnamefont
  {Tsytovich}}\ and\ \bibinfo {author} {\bibfnamefont {I.~M.}\ \bibnamefont
  {Oiringel}},\ }\href@noop {} {\emph {\bibinfo {title} {Polarization
  Bremsstrahlung}}}\ (\bibinfo  {publisher} {Plenum Press},\ \bibinfo {year}
  {1992})\BibitemShut {NoStop}%
\bibitem [{\citenamefont {Tsytovich}(1995)}]{Tsytovich1995}%
  \BibitemOpen
  \bibfield  {author} {\bibinfo {author} {\bibfnamefont {V.~N.}\ \bibnamefont
  {Tsytovich}},\ }\href {http://stacks.iop.org/1063-7869/38/i=1/a=A03}
  {\bibfield  {journal} {\bibinfo  {journal} {Phys. Usp.}\ }\textbf {\bibinfo
  {volume} {38}},\ \bibinfo {pages} {87} (\bibinfo {year} {1995})}\BibitemShut
  {NoStop}%
\bibitem [{\citenamefont {Tsytovich}\ \emph {et~al.}(1996)\citenamefont
  {Tsytovich}, \citenamefont {Bingham}, \citenamefont {De~Angelis},\ and\
  \citenamefont {Forlani}}]{Tsytovich1996}%
  \BibitemOpen
  \bibfield  {author} {\bibinfo {author} {\bibfnamefont {V.~N.}\ \bibnamefont
  {Tsytovich}}, \bibinfo {author} {\bibfnamefont {R.}~\bibnamefont {Bingham}},
  \bibinfo {author} {\bibfnamefont {U.}~\bibnamefont {De~Angelis}}, \ and\
  \bibinfo {author} {\bibfnamefont {A.}~\bibnamefont {Forlani}},\ }\href
  {\doibase 10.1017/S0022377800019140} {\bibfield  {journal} {\bibinfo
  {journal} {J. Plasma Phys.}\ }\textbf {\bibinfo {volume} {56}},\ \bibinfo
  {pages} {127} (\bibinfo {year} {1996})}\BibitemShut {NoStop}%
\bibitem [{\citenamefont {Jackson}(1999)}]{Jackson1999}%
  \BibitemOpen
  \bibfield  {author} {\bibinfo {author} {\bibfnamefont {J.~D.}\ \bibnamefont
  {Jackson}},\ }\href@noop {} {\emph {\bibinfo {title} {Classical
  electrodynamics}}},\ \bibinfo {edition} {3rd}\ ed.\ (\bibinfo  {publisher}
  {Wiley},\ \bibinfo {address} {New York, {NY}},\ \bibinfo {year}
  {1999})\BibitemShut {NoStop}%
\bibitem [{\citenamefont {Diver}(2001)}]{Diver2001}%
  \BibitemOpen
  \bibfield  {author} {\bibinfo {author} {\bibfnamefont {D.~A.}\ \bibnamefont
  {Diver}},\ }\href@noop {} {\emph {\bibinfo {title} {A Plasma Formulary for
  Physics, Technology and Astrophysics}}}\ (\bibinfo  {publisher} {Wiley},\
  \bibinfo {year} {2001})\BibitemShut {NoStop}%
\bibitem [{\citenamefont {Fidone}\ \emph {et~al.}(1988)\citenamefont {Fidone},
  \citenamefont {Granata},\ and\ \citenamefont
  {Johner}}]{FidoneGranataJohner1988}%
  \BibitemOpen
  \bibfield  {author} {\bibinfo {author} {\bibfnamefont {I.}~\bibnamefont
  {Fidone}}, \bibinfo {author} {\bibfnamefont {G.}~\bibnamefont {Granata}}, \
  and\ \bibinfo {author} {\bibfnamefont {J.}~\bibnamefont {Johner}},\ }\href
  {\doibase http://dx.doi.org/10.1063/1.866630} {\bibfield  {journal} {\bibinfo
   {journal} {Phys. Fluids}\ }\textbf {\bibinfo {volume} {31}},\ \bibinfo
  {pages} {2300} (\bibinfo {year} {1988})}\BibitemShut {NoStop}%
\bibitem [{\citenamefont {Karney}\ and\ \citenamefont
  {Fisch}(1985)}]{KarneyFisch1985}%
  \BibitemOpen
  \bibfield  {author} {\bibinfo {author} {\bibfnamefont {C.~F.~F.}\
  \bibnamefont {Karney}}\ and\ \bibinfo {author} {\bibfnamefont {N.~J.}\
  \bibnamefont {Fisch}},\ }\href {\doibase 10.1063/1.865191} {\bibfield
  {journal} {\bibinfo  {journal} {Phys. Fluids}\ }\textbf {\bibinfo {volume}
  {28}},\ \bibinfo {pages} {116} (\bibinfo {year} {1985})}\BibitemShut
  {NoStop}%
\bibitem [{\citenamefont {Munirov}\ and\ \citenamefont
  {Fisch}(2017)}]{Munirov2017a}%
  \BibitemOpen
  \bibfield  {author} {\bibinfo {author} {\bibfnamefont {V.~R.}\ \bibnamefont
  {Munirov}}\ and\ \bibinfo {author} {\bibfnamefont {N.~J.}\ \bibnamefont
  {Fisch}},\ }\href {\doibase 10.1103/PhysRevE.95.013205} {\bibfield  {journal}
  {\bibinfo  {journal} {Phys. Rev. E}\ }\textbf {\bibinfo {volume} {95}},\
  \bibinfo {pages} {013205} (\bibinfo {year} {2017})}\BibitemShut {NoStop}%
\bibitem [{\citenamefont {Widrow}(2002)}]{Widrow2002}%
  \BibitemOpen
  \bibfield  {author} {\bibinfo {author} {\bibfnamefont {L.~M.}\ \bibnamefont
  {Widrow}},\ }\href {\doibase 10.1103/RevModPhys.74.775} {\bibfield  {journal}
  {\bibinfo  {journal} {Rev. Mod. Phys.}\ }\textbf {\bibinfo {volume} {74}},\
  \bibinfo {pages} {775} (\bibinfo {year} {2002})}\BibitemShut {NoStop}%
\bibitem [{\citenamefont {Durrive}\ \emph {et~al.}(2017)\citenamefont
  {Durrive}, \citenamefont {Tashiro}, \citenamefont {Langer},\ and\
  \citenamefont {Sugiyama}}]{Durrive2017}%
  \BibitemOpen
  \bibfield  {author} {\bibinfo {author} {\bibfnamefont {J.-B.}\ \bibnamefont
  {Durrive}}, \bibinfo {author} {\bibfnamefont {H.}~\bibnamefont {Tashiro}},
  \bibinfo {author} {\bibfnamefont {M.}~\bibnamefont {Langer}}, \ and\ \bibinfo
  {author} {\bibfnamefont {N.}~\bibnamefont {Sugiyama}},\ }\href {\doibase
  10.1093/mnras/stx2007} {\bibfield  {journal} {\bibinfo  {journal} {Mon. Not.
  R. Astron. Soc.}\ }\textbf {\bibinfo {volume} {472}},\ \bibinfo {pages}
  {1649} (\bibinfo {year} {2017})}\BibitemShut {NoStop}%
\bibitem [{\citenamefont {Kruer}(2003)}]{Kruer1988}%
  \BibitemOpen
  \bibfield  {author} {\bibinfo {author} {\bibfnamefont {W.~L.}\ \bibnamefont
  {Kruer}},\ }\href@noop {} {\emph {\bibinfo {title} {The Physics of Laser
  Plasma Interactions}}}\ (\bibinfo  {publisher} {Westview Press},\ \bibinfo
  {year} {2003})\BibitemShut {NoStop}%
\end{thebibliography}%

\end{document}